\documentclass[nonacm]{acmart}

\usepackage{color}
\newcommand{\rev}[1]{{\color{black} {#1}}}

\usepackage{graphicx}
\usepackage{makecell}
\def\BibTeX{{\rm B\kern-.05em{\sc i\kern-.025em b}\kern-.08emT\kern-.1667em\lower.7ex\hbox{E}\kern-.125emX}}

\begin{document}

%
\title{Evaluating the Arm Ecosystem for High Performance Computing}

%
\author{Adrian Jackson}
\email{a.jackson@epcc.ed.ac.uk}
\orcid{0000-0003-0073-682X}
\affiliation{%
  \institution{EPCC, The University of Edinburgh}
  \city{Edinburgh, United Kingdom}
}

\author{Andrew Turner}
\email{a.turner@epcc.ed.ac.uk}
\affiliation{%
  \institution{EPCC, The University of Edinburgh}
  \city{Edinburgh, United Kingdom}
}

\author{Mich\`{e}le Weiland}
\email{m.weiland@epcc.ed.ac.uk}
\orcid{0000-0003-4713-3073}
\affiliation{%
  \institution{EPCC, The University of Edinburgh}
  \city{Edinburgh, United Kingdom}
}

\author{Nick Johnson}
\email{n.johnson@epcc.ed.ac.uk}
\affiliation{%
  \institution{EPCC, The University of Edinburgh}
  \city{Edinburgh, United Kingdom}
}

\author{Olly Perks}
\email{olly.perks@arm.com}
\affiliation{%
  \institution{Arm}
  \city{Warwick, United Kingdom}
}

\author{Mark Parsons}
\email{m.parsons@epcc.ed.ac.uk}
\affiliation{%
  \institution{EPCC, The University of Edinburgh}
  \city{Edinburgh, United Kingdom}
}

%

%
\begin{abstract}
In recent years, Arm-based processors have arrived on the HPC scene, offering an alternative the existing status quo, which was largely dominated by x86 processors. In this paper, we evaluate the Arm ecosystem, both the hardware offering and the software stack that is available to users, by benchmarking a production HPC platform that uses Marvell's ThunderX2 processors. We investigate the performance of complex scientific applications across multiple nodes, and we also assess the maturity of the software stack and the ease of use from a users' perspective. This papers finds that the performance across our benchmarking applications is generally as good as, or better, than that of well-established platforms, and we can conclude from our experience that there are no major hurdles that might hinder wider adoption of this ecosystem within the HPC community.
\end{abstract}

%
%


%

\maketitle

\section{Introduction}
There is an established ecosystem of server-level processors suitable for computational simulation and machine learning applications built around traditional x86 architectures from processor manufacturers such as Intel and AMD. However, recently, alternative processor technologies have been developed; foremost amongst these are Arm based processors from manufacturers such as Marvell (ThunderX2), Ampere (eMAG), Huawei (Kunpeng 920), Fujitsu (A64FX) and Amazon (Graviton).

The first server class processor to be commercially available in large volume is the ThunderX2 processor from Marvell. The ThunderX2 processor uses the Armv8 instruction set and it has been designed specifically for server workloads. The design includes eight DDR4 memory channels to deliver measured STREAM triad memory bandwidth in excess of 220 GB/s per dual-socket node.

However, hardware only represents one part of the ecosystem that is required to deliver a usable High Performance Computing (HPC) platform for the varied workloads of computational simulation and machine learning applications. Operating system, compiler, and library support is required to provide a functional environment that supports large scale HPC applications and to ensure applications can both be easily ported to such new hardware as well as efficiently exploit it.

In this paper we will evaluate a range of computational simulation applications on a HPC system comprised of nodes with ThunderX2 processors connected together with an Infiniband network. Our paper makes the following contributions to deepening the understanding of the performance of a production HPC system that is based on the Arm ecosystem:
\begin{enumerate}
    \item We outline performance measurements of the interconnect network, using established MPI benchmarks, allowing us to assess the potential scaling performance of distributed memory applications.
    \item We present and evaluate the multi-node performance of scientific applications with varying performance characteristics and compare it to the established x86 ecosystem.
    \item We evaluate the portability of applications onto this new system, compared with equivalent systems based on other processor technologies.
    \item We discuss the causes for the performance and scalability results that have been observed, and based on this we draw conclusions with regards to the maturity of the current generation of Arm-based systems for HPC.
\end{enumerate}

\section{Related work}
Some initial porting and performance evaluation work on ThunderX2 processors has been documented~\cite{Bristol_CUG18}, outlining performance and cost benefits for some applications using ThunderX2 compared to other available Intel processors. However, these results only consider single node performance, using at most two processors, and do not consider the performance, behaviour, or functionality associated with multi-node applications. The majority of computational simulation applications~\cite{ARCHER_Usage}
on large scale HPC systems require many nodes for simulations. \rev{The Mont-Blanc project has published a detailed energy usage study for a ThunderX2-based system, using specialist on-node hardware to measure the energy to solution of a number of benchmarks and mini-apps \cite{Arm_energy}}. This paper investigates the performance of distributed memory communications (MPI), as well as scientific applications that use MPI, on a ThunderX2 based system, and the associated libraries required for functionality and performance. 

HPC platforms are evaluated using a wide range of benchmarks, each targeting a different performance aspect; popular benchmark suites include~\cite{SPECMPI}~\cite{HPCChallenge}~\cite{NASBenchmark}. These include application specific benchmarks~\cite{ARCHER_Benchmarks}, and have included benchmarking application across multiple systems~\cite{ARCHER_bench_perf_report}.

\section{Benchmarking methodology} 
In order to fully evaluate the performance of the Arm-based HPE Apollo 70, we execute a range of benchmarks and applications that stress different aspects of the architecture and the software stack, and compare our results with well established HPC systems. Our benchmarking methodology adheres to the following principles:

\paragraph{MPI performance} The scalability of multi-node workloads relies on a high-performance interconnect and efficient inter-node communication. As the majority of HPC workloads rely on MPI to distribute work across compute nodes, we use MPI benchmarks to assess the communication performance on the HPE Apollo 70. There are a range of MPI libraries and implementations available on the benchmarking systems we are using.
\paragraph{Reproducibility} We use process and thread pinning to cores to ensure our results are not impacted or skewed by the operating system's process/thread management policies, and are reproducible. We also list the compiler versions and flags, as well as the libraries used, in Table~\ref{tab:comp}.
\paragraph{Applications} The applications chosen for this benchmarking study cover different scientific domains, programming languages, libraries and performance characteristics (i.e. compute intensive, memory-bandwidth limited, etc). They also represent widely used real-life applications. GROMACS and OpenSBLI are part of the UK HPC benchmarks\cite{ARCHER_Benchmarks}. As we are chiefly interested in the compute performance of the applications, we disabled or reduced I/O (in particular output) as much as possible. 
\paragraph{Multi-node benchmarks} The scaling behaviour of the applications is benchmarked and analysed, starting from a single node to up to 32 nodes in total. This allows for the identification of performance bottlenecks caused by the network or the communication libraries.
\paragraph{Performance comparison} The results that were achieved on the HPE Apollo 70 are compared with those from three well-established and widely deployed platforms in order to assess the relative performance of the system. The results are generally compared on a per-node basis (rather than per-core) and the test cases configurations are the same across systems.

\section{Benchmarking systems} 

The system under evaluation, an Arm-based HPE Apollo 70, is compared against well established HPC system architectures with very mature software ecosystems. This not only enables us to compare performance due to difference in the hardware, but also allows us to assess the ease of porting applications and to evaluate the relative maturity of the software stack that is available on the Arm system. Details on each of the systems used for this benchmarking activity are given below, and Table \ref{tab:node} summarises their compute node specifications. \rev{We have also include a calculated node memory bandwidth result for each system, collected using the STREAM benchmark, to allow comparison of the achievable memory bandwidth between the different hardware solutions.}

\begin{table*}[t]
\caption{Compute node specifications.}
\label{tab:node}
\begin{tabular}{|r|c|c|c|c|}
\hline
 & \bf{HPE Apollo 70} & \bf{SGI ICE XA} & \bf{Cray XC30} & \bf{Dell EMC} \\
\hline
\hline
Processor & \makecell{Marvell \\ ThunderX2} & \makecell{Intel Xeon \\ E5-2695} & \makecell{Intel Xeon\\ E5-2697 v2} & \makecell{Intel Xeon Gold \\ 6142}  \\
 & {(ARMv8)} & {(Broadwell)} & {(IvyBridge)} & {(Skylake)} \\
\hline
Processor clock speed & 2.2GHz & 2.1GHz & 2.7GHz & 2.6GHz \\
Cores per processor & 32 & 18 & 12 & 16  \\
Cores per node & 64 & 36 & 24 & 32  \\
Threads per core & \textbf{1}, 2 or 4 & \textbf{1} or 2 & \textbf{1} or 2 & 1 \\
Vector width & 128bit & 256bit & 256bit & 512bit \\
\hline
Double Precision (DP) FLOPS per cycle & 8 & 16 & 8 & 32 \\
Maximum node DP GFLOP/s & 1126.4 & 1209.6 & 518.4 & 2662.4 \\
\hline
L1 cache (core) & 32KB  & 32KB & 32KB & 32KB  \\
L2 cache (core) & 256KB & 256KB & 256KB & 1024KB  \\
L3 cache (shared) & 32MB & 45MB & 30MB & 22MB  \\
\hline
Memory per node & 256GB & 256GB & \textbf{64GB}/128GB & \textbf{192GB}/384GB  \\
Memory per core & 4GB & \rev{7.11GB} & \textbf{2.66GB}/5.33GB & \textbf{6GB}/12GB   \\
Memory channels (processor) & 8 & 4 & 4 & 6   \\
Specified memory bandwidth (processor) & 160GB/s & \rev{76.8GB/s} &  \rev{59.7GB/s}  & 119.2GB/s \\
\rev{Specified memory bandwidth (core)} & 5.0GB/s & \rev{3.3GB/s} & 6.4GB/s & \rev{7.5GB/s} \\
\rev{Measured memory bandwidth (node)} & \rev{221.48GB/s} & \rev{97.85GB/s} & \rev{72.86GB/s} & \rev{144.26GB/s} \\
\hline
\end{tabular}
\end{table*}

\paragraph{HPE Apollo 70} The Arm-based system under evaluation is an HPE Apollo 70 cluster with dual-socket compute nodes connected with Mellanox EDR Infiniband (IB) using a non-blocking fat tree topology. Each compute node consists of two 32-core Marvell ThunderX2 processors running at 2.2GHz, and 256GB DDR4 memory. The processors on this system are set to use 1 hardware thread (SMT) per core.

\paragraph{SGI ICE XA} The SGI ICE XA system has compute nodes each with two 2.1 GHz, 18-core, Intel Xeon E5-2695 (Broadwell) series processors. They have 256 GB of memory shared between the two processors. The system has a single IB FDR fabric with a bandwidth of 54.5 Gb/s.

\paragraph{Cray XC30} This system has 24 cores (two Intel Xeon 2.7 GHz, 12-core E5-2697v2 processors) and 64 GB of DDR3 memory per node (128 GB on a small number of large memory nodes). Nodes are connected by the Cray Aries network.

\paragraph{Dell EMC} The Dell EMC PowerEdge system has two 16-core Intel Xeon Gold Skylake processors per compute node, running at 2.6GHz, and with up to 384 GB of DDR4 memory shared between to processors. The system uses Intel's OmniPath interconnect.

\section{MPI Performance}

We have used the Intel MPI Benchmark suite to compare the MPI performance on the HPE Apollo 70 with that of the other systems evaluated, including the  Cray XC30, which is known to demonstrate excellent interconnect performance. \rev{We present both point-to-point and collective performance, focussing on the $MPI\_Sendrecv$ benchmark for point-to-point message and the $MPI\_Alltoallv$ and $MPI\_Allgatherv$ collectives. We focus on these collectives specifically as they are generally expensive operations that are important to a range of HPC applications.}  We report results for the Alltoallv and Allgatherv benchmarks for two message sizes: 128 KiB and 1 MiB. \rev{The Sendrecv benchmark was run in the on-cache mode, meaning MPI processes on the same processor can utilise the last level cache for communicating data.}

For each of the systems we used the following MPI libraries:
\begin{itemize}
    \item HPE Apollo 70: HPE MPT 2.20
    \item HPE Apollo 70: OpenMPI 4.0.0
    \item \rev{SGI ICE XA: SGI MPT 2.16}
    \item Cray XC30: Cray MPICH2 7.5.5
    \item \rev{Dell EMC: Intel MPI 2017.4}
\end{itemize}

Tables~\ref{tab:alltoallv_128K} and~\ref{tab:alltoallv_1M} show the Alltoallv test results for the 128KiB and 1MiB message sizes respectively. Tables~\ref{tab:allgatherv_128K} and~\ref{tab:allgatherv_1M} show the Allgatherv test results for the 128KiB and 1MiB message sizes.

\begin{table*}[h]
\caption{Mean Alltoallv Intel MPI Benchmark test time (ms) for a message size of 128 KiB as a function of node count.}
\label{tab:alltoallv_128K}
\begin{tabular}{|c|r|r|r|r|r|}
\hline
\bf{Nodes} & \bf{HPE Apollo 70 (MPT)} & \bf{HPE Apollo 70 (OpenMPI)} & \bf{SGI ICE XA} & \bf{Cray XC30} & \bf{Dell EMC} \\
\hline
\hline
2 & 99.531 & 105.189 & 74.264 & 11.731 & 10.852  \\
4 & 385.304 & 322.071 & 142.151 & 37.512 & 36.380  \\
8 & 887.104 & 784.300 & 283.167 & 117.046 & 86.948 \\
16 & 1,863.269  & 1,679.783 & 692.317 & 293.256 & 278.481 \\
\hline
\end{tabular}
\end{table*}

\begin{table*}[h]
\caption{Mean Alltoallv Intel MPI Benchmark test time (ms) for a message size of 1 MiB as a function of node count.}
\label{tab:alltoallv_1M}
\begin{tabular}{|c|r|r|r|r|r|}
\hline
\bf{Nodes} & \bf{HPE Apollo 70 (MPT)} & \bf{HPE Apollo 70 (OpenMPI)} & \bf{SGI ICE XA} & \bf{Cray XC30} & \bf{Dell EMC} \\
\hline
\hline
2 & 754.835 & 833.083 & 231.816 & 95.763 & 89.888 \\
4 & 2,966.812 & 2,583.005 & 2,135.086 & 307.588 & 307.026 \\
8 & 6,971.923 & 6,086.923 & 14,811.546 & 977.187 & 783.750 \\
16 & 15,544.344 & 13,065.089 & Out of Memory & 2,263.569 &  2,237.830 \\
\hline
\end{tabular}
\end{table*}

\begin{table*}[h]
\caption{Mean Allgatherv Intel MPI Benchmark test time (ms) for a message size of 128 KiB as a function of node count.}
\label{tab:allgatherv_128K}
\begin{tabular}{|c|r|r|r|r|r|}
\hline
\bf{Nodes} & \bf{HPE Apollo 70 (MPT)} & \bf{HPE Apollo 70 (OpenMPI)} & \bf{SGI ICE XA} & \bf{Cray XC30} & \bf{Dell EMC} \\
\hline
\hline
2 & 18.840 & 10.061 & 29.106 & 4.016 & 7.452 \\
4 & 39.174 & 18.182 & 16.394 & 7.893 & 14.658 \\
8 & 86.769 & 36.039 & 30.127 & 19.706 & 29.214 \\
16 & 190.696 & 70.600 & 70.542 & 44.222 & 60.268 \\
\hline
\end{tabular}
\end{table*}

\begin{table*}[h]
\caption{Mean Allgatherv Intel MPI Benchmark test time (ms) for a message size of 1 MiB as a function of node count.}
\label{tab:allgatherv_1M}
\begin{tabular}{|c|r|r|r|r|r|}
\hline
\bf{Nodes} & \bf{HPE Apollo 70 (MPT)} & \bf{HPE Apollo 70 (OpenMPI)} & \bf{SGI ICE XA} & \bf{Cray XC30} & \bf{Dell EMC} \\
\hline
\hline
2 & 163.610 & 101.678 & 59.093 & 32.400 & 50.102 \\
4 & 339.137 & 190.665 & 119.589 & 65.496 & 93.327 \\
8 & 740.027 & 352.193 & 237.253 & 147.087 & 188.210 \\
16 & 1,611.643 & 689.009 & 552.725 & 297.995 & 375.877 \\
\hline
\end{tabular}
\end{table*}

The results from the MPI benchmarks highlight that the performance of some collective operations on the HPE Apollo 70 system are poor when compared to the Cray XC30. Initial investigations point towards possible configuration issues with the Infiniband interconnect and MPI library, and work is ongoing to resolve this situation, as poor performance on MPI collectives will have a negative impact on the scalability of applications that heavily rely on these operations.

\rev{We can see from the point-to-point performance, outlined in Table~\ref{tab:sendrecv}, that the Apollo 70 system has similar performance to the SGI ICE XA system, both using Infiniband networks. It is interesting that the MPT library on the Apollo 70 system achieves significantly higher bandwidth and lower latency for the on-node communication benchmarks when compared to OpenMPI. It is likely that the MPT library is taking advantage of communicating through the cache, an optimisation that does not appear to be enabled for OpenMPI, but which should be configurable to enable achieving similar performance for OpenMPI. It is also worth noting that this benchmark was using only 2 MPI processes on the node, providing an uncontested environment for the benchmark. It is likely different performance would be observed if larger numbers of MPI processes were occupying the node. We also see that the Cray XC30 exhibits better performance that the other, Infiniband or Omnipath-based, systems.}

\begin{table*}[h]
\caption{Mean Sendrecv Intel MPI Benchmark test time (ms) and bandwidth (MB/s)}
\label{tab:sendrecv}
\begin{tabular}{|c|r|r|r|r|r|}
\hline
& \multicolumn{5}{c|}{\makecell{Mean test time (ms) \it{(Bandwidth GB/s)}}} \\
\hline
\bf{Configuration} & \makecell{\bf{HPE Apollo 70} \\\bf{(MPT)}} & \makecell{\bf{HPE Apollo 70} \\ \bf{(OpenMPI)}} & \bf{SGI ICE XA} & \bf{Cray XC30} & \bf{Dell EMC} \\
\hline
\hline
\makecell{within a node \\ same socket} & 0.46 \it{(21,114.09)} & 0.33 \it{(13,655.82)} & 0.58 \it{(21,342.88)} & 0.38 \it{(26,196.39)}  & 0.42 \it{(18,796.07)} \\
\hline
\makecell{within a node \\ different sockets} & 0.75 \it{(18,751.15)} & 0.70 \it{(12,402.8)} & 0.78 \it{(21,346.13)} & 0.69 \it{(26,329.70)}  & 0.87 \it{(18,902.10)} \\
\hline
\makecell{between nodes} & 2.17 \it{(10,646.78)} & 2.04
 \it{(11,082.26)}  & 1.67 \it{(12,517.62)} & 1.42 \it{(14,506.03)}  & 1.52 \it{(17,886.42)} \\
\hline
\end{tabular}
\end{table*}

\section{Overview of benchmarking applications}

To evaluate the performance of the Apollo 70 system we have chosen a set of application and benchmarks that have a range of different performance characteristics, functionalities, and implementations.

We have chosen codes that span the main programming languages typically used within parallel computing (i.e.\ C, C++, and Fortran) and that have MPI, OpenMP, and hybrid (for our chosen applications MPI and OpenMP) parallel implementations.

We have also ensured that we have a range of different performance characteristics in the applications, from memory bandwidth bound codes, to those dependent on the network for  good performance when using multiple nodes. We describe each application in turn in the following paragraphs, together with the test cases and the configuration used for benchmarking.

\subsection{HPCG}

High Performance Conjugate Gradients (HPCG~\cite{HPCG}) is a benchmark for large-scale parallel systems that aims to represent the computational and data access patterns of a broad set of important computational simulation applications. The conjugate gradients algorithm used in the benchmark is not just floating point performance limited, it is also heavily reliant on the performance of the memory system, and to a lesser extend on the network used to connect the processors together. The following functionality is implemented in benchmark:
\begin{itemize}
    \item Sparse matrix-vector multiplication.
    \item Vector updates.
    \item Global dot products.
    \item Local symmetric Gauss-Seidel smoother.
    \item Sparse triangular solve (as part of the Gauss-Seidel smoother).
    \item Multigrid pre-conditioned solvers.
\end{itemize}

This functionality helps to ensure it must provide a performance assessment that is more comparable to the performance that real applications could achieve for a given system, compared to other benchmarking kernels used for assessing HPC systems, such as High Performance LINPACK (HPL\cite{HPL}). For this reason, \rev{HPCG is rapidly becoming the second most important benchmark for evaluating HPC system performance, and is now included in the Top500 list}\footnote{https://www.top500.org/hpcg/lists/2018/11/}.

There is a reference implementation of HPCG available, written in C++ with MPI and OpenMP support. However, there are also optimised versions of HPCG available for a range of hardware, including Intel processors \footnote{https://software.intel.com/en-us/mkl-linux-developer-guide-overview-of-the-intel-optimized-hpcg} and Arm systems \footnote{https://developer.arm.com/products/software-development-tools/hpc/resources/porting-and-tuning/building-hpcg-with-arm-compiler}. These generally provide improved performance over the reference implementation. For this study we used the optimised versions of HPCG.

\subsubsection{Test Case}
The HPCG benchmark can be configured to run different size problems by providing the size of the local data block each process works on. To ensure consistency across the systems the same local block size was used for all the benchmarks: \texttt{--nx=64 --ny=64 --nz=64}. This size of local domain block was used because the optimised Arm version of HPCG is restricted to using sizes that are divisible by 2 at all grid scales. This block size may not give the best performance across all systems, but it does provide a consistent benchmark.

\subsubsection{Configuration}
We are using version 3.0 of HPCG for these benchmarking. For all systems a range of pure MPI and hybrid (MPI + OpenMP) configurations were tested to evaluate which gave the best performance. We used the optimal configuration for each system to run the benchmarks, outlined in Table~\ref{tab:hpcg_config}.

\begin{table}[h]
\caption{HPCG process and thread configuration}\label{tab:hpcg_config}
\begin{tabular}{|c|c|c|}
\hline
\bf{System} & \bf{\makecell[l]{MPI processes \\ per node}} & \bf{\makecell[l]{OpenMP threads \\ per MPI process}}  \\
\hline
\hline
HPE Apollo 70 & 32 & 2  \\
SGI ICE XA & 18 & 2  \\
Cray XC30 & 12 & 2 \\
Dell EMC & 32 & 1  \\
\hline
\end{tabular}
\end{table}








\begin{table*}[t]
\caption{Compilers, Compiler Flags and Libraries.}\label{tab:comp}
\begin{tabular}{|r|l|l|l|}
\hline
 & \bf{Compiler} & \bf{Compiler flags} & \bf{Libraries} \\
\hline
\hline
\bf{COSA} &  &  &  \\
\hline
HPE Apollo 70 & GNU Fortran v8.2 & \makecell[l]{-g -fdefault-double-8 -fdefault-real-8 -fcray-pointer \\ -ftree-vectorize -O3  -ffixed-line-length-132} & \makecell[l]{HPE MPT MPI library (v2.20) \\ ARM Performance Libraries (v19.0.0)} \\
\hline
SGI ICE XA & GNU Fortran v8.2 & 
\makecell[l]{-g -fdefault-double-8 -fdefault-real-8 -fcray-pointer \\ -ftree-vectorize -O3  -ffixed-line-length-132} & \makecell[l]{SGI MPT 2.16 \\ Intel MKL 17.0.2.174} \\
\hline
Cray XC30 & GNU Fortran v7.2 & \makecell[l]{-g -fdefault-double-8 -fdefault-real-8 -fcray-pointer \\ -ftree-vectorize -O3  -ffixed-line-length-132} & \makecell[l]{Cray MPI library (v7.5.5) \\ Cray LibSci (v16.11.1)} \\
\hline
\hline
\bf{GROMACS} &  &  &  \\
\hline
HPE Apollo 70 & Arm Clang 19.0.0 & \makecell[l]{ -std=c++11 -mcpu=native -O3 -DNDEBUG -funroll-all-loops \\ -fexcess-precision=fast -fPIC -O3} & \makecell[l]{OpenMPI 4.0.0 \\ FFTW 3.3.8} \\
\hline
SGI ICE XA & GCC 6.2.0 & \makecell[l]{-march=core-avx2 -std=c++0x -O3 -funroll-all-loops \\ -fexcess-precision=fast } & \makecell[l]{SGI MPT 2.16 \\ FFTW 3.3.5} \\
\hline
Cray XC30 & GCC 6.3.0 & \makecell[l]{-mavx   -static -O3 -ftree-vectorize -funroll-loops -std=c++11 \\ -O3 -funroll-all-loops -fexcess-precision=fast   } & \makecell[l]{Cray MPICH2 7.5.5 \\ FFTW 3.3.6} \\
\hline
Dell EMC & Intel 17.4 & \makecell[l]{-xCORE-AVX512 -mkl=sequential  -std=gnu99 -std=c++11 \\ -O3 -ip -funroll-all-loops -alias-const -ansi-alias \\ -no-prec-div -fimf-domain-exclusion=14 -qoverride-limits} & \makecell[l]{Intel MPI library 17.4 \\ Intel MKL 17.4} \\
\hline
\hline
\bf{HPCG} &  &  &  \\
\hline
HPE Apollo 70 & Arm Clang 19.0.0 & \makecell[l]{ -O3 -ffast-math -funroll-loops -fopenmp \\ -std=c++11 -ffp-contract=fast -mcpu=native} & \makecell[l]{OpenMPI 4.0.0 \\ ARM Performance Libraries (v19.0.0)} \\
\hline
SGI ICE XA & Intel 17.0.2.174  & \makecell[l]{-xCORE-AVX2 -qopenmp -std=c++11 \\ -O3 -DNDEBUG} & \makecell[l]{SGI MPT 2.16 \\ Intel MKL 17.0.2.174 } \\
\hline
Cray XC30 & Intel 17.0.0.098 & \makecell[l]{-qopenmp  -std=c++0x -O3 -DNDEBUG} & \makecell[l]{Cray MPICH2 7.5.5 \\ Intel MKL 17.0.0.098 } \\
\hline
\hline
\bf{OpenSBLI} &  &  &  \\
\hline
HPE Apollo 70 & Arm Clang 19.0.0 & \makecell[l]{-O3 -std=c99 -fPIC -Wall} & \makecell[l]{OpenMPI 4.0.0 \\ HDF5 1.10.4} \\
\hline
SGI ICE XA & Intel 17.0.2.174 & \makecell[l]{\rev{-O3 -ipo -restrict -fno-alias }} & \makecell[l]{SGI MPT 2.16 \\ HDF5 1.10.1} \\
\hline
Cray XC30 & Cray Compiler v8.5.8 & \makecell[l]{-O3 -hgnu} & \makecell[l]{Cray MPICH2 (v7.5.2) \\ HDF5 (v1.10.0.1)} \\
\hline
Dell EMC & Intel 17.4 & \makecell[l]{\rev{-O3 -ipo -restrict -fno-alias }} & \makecell[l]{Intel MPI 17.4 \\ HDF5 1.10.1} \\
\hline
\hline
\bf{Nektar++} &  &  &  \\
\hline
HPE Apollo 70 & GNU 8.2.0 & \makecell[l]{-O3 -fPIC -DNDEBUG} & \makecell[l]{OpenMPI 4.0.0 \\ ARM Performance Libraries (v19.0.0) \\ FFTW 3.2.2 \\ Boost 1.6.7}
 \\
 \hline
\rev{HPE Apollo 70} & \rev{ARM Clang 19.0.0} & \makecell[l]{-O3 -fPIC -DNDEBUG} & \makecell[l]{OpenMPI 4.0.0 \\ ARM Performance Libraries (v19.0.0) \\ FFTW 3.2.2 \\ Boost 1.6.7}
\\
\hline
SGI ICE XA & GCC 6.2.0  & \makecell[l]{-O3 -fPIC -DNDEBUG} & \makecell[l]{SGI MPT 2.18 \\ FFTW 3.2.2 \\ Boost 1.6.7 } \\
\hline
\end{tabular}
\end{table*}

\subsection{COSA}
COSA supports steady, time-domain (TD), and frequency-domain (harmonic balance or HB) solvers, implementing the numerical solution of the Navier-Stokes (NS) equations using a finite volume space-discretisation and multigrid (MG) integration.  It is implemented in Fortran~\cite{COSA_CSRD} and has been parallelised using MPI~\cite{MPI}, with each MPI process working on a set of grid blocks (geometric partitions) of the simulation.  

In the HB solver there exists an additional dimension compared to the steady and TD solvers, which can be viewed as a harmonic varying from 1 to Nh, a user specified number of elemental flow harmonics. However, the code does not solve directly such elemental harmonics, but rather Nh equally time-spaced snapshots of the required periodic flow field, linked to the Nh elemental harmonics using a Fourier transform.

The code is structured so that the core computational kernels can, for the most part, be reused for the steady solvers and HB simulations, with HB simply requiring an outer loop over the Nh snapshots using the steady solver kernels. COSA has been shown to exhibit good parallel scaling to large numbers of MPI processes with a sufficiently large test case~\cite{COSA_CF}.

\subsubsection{Test Case} \label{sec:cosa_test_case}
For the benchmarking of COSA we used a HB test case with 4 harmonics and a grid composed of $16,384$ blocks, encompassing a total of $47,071,232$ grid cells. This limits the MPI domain decomposition to, at most, $16,384$ MPI processes. This test case requires a minimum of $1$ Terabyte (TB) of memory across the MPI processes running the simulation, meaning it is not possible to run using a single process or node from the test systems we had access to. To enable fast benchmarking the simulation was run for 20 iterations, which is a significantly smaller number of iteration than a production run would typically use but is enough to provide sufficient simulation functionality to evaluate performance.

\subsubsection{Configuration}

Writing output data to storage can be a significant overhead in COSA, especially for simulations using small numbers of iterations. To ensure variations in the I/O hardware of the platforms being benchmarked does not affect the results collected we disabled output functionality for benchmark runs of COSA.

The benchmark was run with a single MPI process per core, and all the cores in the node utilised for the benchmarks (i.e. we were not under-populating nodes except for the case of the Cray XC30 using 512 MPI processes, as this left one node with 16 free cores).

\subsection{GROMACS}

GROMACS \cite{GROMACS} is a versatile package to perform molecular dynamics, i.e. simulate the Newtonian equations of motion for systems with hundreds to millions of particles.

It is primarily designed for biochemical molecules like proteins, lipids and nucleic acids that have a lot of complicated bonded interactions, but since GROMACS is extremely fast at calculating the non-bonded interactions (that usually dominate simulations) many groups are also using it for research on non-biological systems, e.g. polymers.

\subsubsection{Test Case}

The GROMACS 1400k atom benchmark is taken from the HEC BioSim website \cite{HECBioSim_benchmarks} and has also been used within the ARCHER benchmarks with results freely-available online \cite{ARCHER_Benchmarks}. This strong-scaling benchmark represents a Pair of hEGFR Dimers of 1IVO and 1NQL and we would expect this benchmark to scale well up to 16-32 compute nodes (depending on the performance of the MPI library and interconnect). This benchmark case performs minimal I/O.

\subsubsection{Configuration}

The single-precision version of GROMACS was used for all benchmark runs.  The GROMACS \texttt{mdrun} command line option \texttt{-noconfout} was specified for all benchmark runs. Performance of the GROMACS benchmark is measured in ns/day read directly from the GROMACS output log file.

For all systems, 1 MPI process was used per physical core with 1 OpenMP thread per MPI process. A single hardware thread was used in all cases. All compute node cores were used in all benchmark runs (i.e. all compute nodes were fully populated).

\subsection{OpenSBLI}

OpenSBLI \cite{OSBLI_2017} is a Python-based modelling framework that is capable of expanding a set of differential equations written in Einstein notation, and automatically generating C code that performs the finite difference approximation to obtain a solution. This C code is then targeted with the OPS library towards specific hardware backends, such as MPI/OpenMP for execution on CPUs, and CUDA/OpenCL for execution on GPUs.

The main focus of OpenSBLI is on the solution of the compressible Navier-Stokes equations with application to shock-boundary layer interactions (SBLI). However, in principle, any set of equations that can be written in Einstein notation may be solved.

\subsubsection{Test Case}

The benchmark test case setup using OpenSBLI is the Taylor-Green vortex problem in a cubic domain of length $2\pi$. For this study, we have investigated the strong scaling properties for the benchmark on grids of sizes $512\times512\times512$ and $1024\times1024\times1024$. The larger benchmark should be more demanding in terms of both computation and memory access. The smaller benchmark is included to allow comparisons between single nodes of different architectures as the larger benchmark requires a minimum of 2 compute nodes. This benchmark performs minimal I/O.

\subsubsection{Configuration, Compilation and Libraries}

Performance is measured in 'iterations/s'. The total runtime and number of iterations are read directly from the OpenSBLI output and these are used to compute the number of iterations per second.

For all systems, 1 MPI process was used per physical core with 1 OpenMP thread per MPI process. A single hardware thread was used in all cases.

\subsection{Nektar++}

Nektar++~\cite{NektarPaper} is a tensor-product based finite-element framework designed to allow construction of efficient, classical, low polynomial order h-type solvers (where h is the size of the finite-element) as well as higher p-order piece-wise polynomial order solvers. Written in C++, Nektar++ is run here as pure-MPI code but can run in a hybrid MPI-OpenMP mode, can make use of hyperthreading and supports use of PGAS communications models via extensions to it's communications library. It is under active development by Imperial College London and the University of Utah.

\subsubsection{Test Case}

The geometry used for this test case is a 3D segment of a rabbit's descending aorta with two pairs of intercostal arteries branching off. The inlet has a diameter $D = 3.32mm$. The solver used is the incompressible Navier Stokes solver and run for 4000 timesteps of 1 $\mu$s giving 4 ms simulated runtime with a backward Euler time integration method. The output is limited to a single write of the flow fields at the end of the simulation.

\subsubsection{Configuration, Compilation and Libraries}

Performance is measured from the code's "time integration" output which maps to the time spent running the solver routines for the test case. For both systems tested, a single MPI process per physical core was used without shared memory processing; this was run as a purely MPI code to limit the differences between systems.

\section{Results and Evaluation}
This section presents and evaluates the benchmarking applications on a range of node counts to investigate both performance and scalability across the different systems. For \rev{HPCG, COSA, OpenSBLI and GROMACS, the HPE Apollo 70 results are contrasted with all three comparison platforms. Nektar++ is compared to the SGI ICE XA only}. At the end of this section, we present our evaluation of portability and the Arm software stack maturity.

\subsection{HPCG}
HPCG is considered to be a more representative HPC benchmark than HPL as it has a more complex resource usage pattern, more akin to real HPC applications. As such, HPCG performance will take into account memory bandwidth, floating point performance and to some extent network performance.

\begin{table}[h]
\caption{Single node HPCG GFlop/s.}\label{tab:hpcg_results_single_node}
\begin{tabular}{|c|r|}
\hline
\bf{System} & \bf{Performance} \\
\hline
\hline
HPE Apollo 70 & 30.529  \\
SGI ICE XA & 21.115  \\
Cray XC30 & 15.650 \\
Dell EMC (AVX512) & 34.581  \\
Dell EMC (AVX2) & 28.120 \\
\hline
\end{tabular}
\end{table}

Table~\ref{tab:hpcg_results_single_node} shows the HPCG performance for a single node \rev{across the test systems. The Dell EMC system was tested with different compilation options targeting the wide vectorisation functionality. The table demonstrates that a significant performance benefit ($\sim20\%$) is achieved on the Skylake processors in the Dell EMC system when specifically targeting the AVX-512 instruction set. The table also demonstrates that the Apollo 70 nodes are performing very well}, providing around $2.7\%$ of theoretical total GFlop/s of a node compared to \rev{$1.3\%$ for the Dell EMC system (with Intel Skylake processors) and} $1.7\%$ for the ICE XA system (with Intel Broadwell processors). The XC30 system has slightly better single node performance ($3.0\%$ of theoretical GFlop/s), albeit with a much lower total floating point capability than the \rev{other systems compared}.

\begin{table}[h]
\caption{\rev{Multi-node HPCG GFlop/s.}}\label{tab:hpcg_results_multi_node}
\begin{tabular}{|c|r|r|r|r|}
\hline
\bf{Nodes} & \makecell{\bf{HPE} \\ \bf{Apollo 70}} & \makecell{\bf{SGI} \\ \bf{ICE XA}} & \bf{Cray XC30} & \makecell{\bf{Dell EMC} \\\bf{(AVX2)}} \\
\hline
\hline
2 & 61.185 &  40.491 & 32.048 & 56.505 \\
4 & 120.972 & 80.317 & 61.617 & 111.598 \\
8 & 241.363 & 158.774 & 120.749 & 214.358 \\
16 & 482.202 & 305.534  & 245.870 & 400.320 \\
32 & 939.109 & 602.092 & 491.101 & 747.066 \\
\hline
\end{tabular}
\end{table}

\rev{Table~\ref{tab:hpcg_results_multi_node} presents HPCG results when using multiple nodes. Problems with running the AVX-512 version of the benchmark on the Dell EMC system have restricted us to providing only the AVX2 numbers}. We see that the Apollo 70 maintains nearly twice the total GFlop/s when compared to the Cray XC30 across a range of node counts, \rev{and outperforms both the SGI ICE XA and Dell EMC system. If we were to assume linear scaling for the AVX-512 results from a single node up to 32 nodes on the Dell EMC system, this translates to around $10\%$ more GFlop/s than the Apollo 70.} 

\rev{We undertook a Pearson correlation analysis of HPCG performance on the various systems against the theoretical peak floating point performance per node, the measured memory bandwidth per node (both detailed in Table~\ref{tab:node}), the inter-node Sendrecv latency and bandwidth (both detailed in Table~\ref{tab:sendrecv}). The Pearson correlation analysis assesses the level of correlation between the values from two datasets, reporting correlation that varies between -1 (absolute negative correlation) and +1 (absolute positive correlation). Whilst we only have a limited number of data points in our datasets, the correlation analysis does provide some indications of the impact of various performance characteristics of the various systems. The results of this analysis, detailed in Table~\ref{tab:hpcg_correl}, demonstrate the impact of aggregate memory bandwidth on multi-node HPCG performance, and the importance of floating point capability for the single node results. It is also noted that MPI message latency seems to have a reasonable correlation with the multi-node HPCG performance.}

\begin{table}[h]
\caption{\rev{HPCG Correlation values with theoretical peak GFlop/s per node, measured memory bandwidth per node, off-node MPI latency and bandwidth.}}\label{tab:hpcg_correl}
\begin{tabular}{|c|r|r|r|r|}
\hline
\bf{Nodes} & \makecell{\bf{Node DP}\\ \bf{GFlop/s}} & \makecell{\bf{Memory} \\ \bf{bandwidth}} & \makecell{\bf{MPI}\\ \bf{latency}} & \makecell{\bf{MPI}\\ \bf{bandwidth}} \\
\hline
\hline
1 & 0.83 &  0.76 & 0.42	& 0.25 \\
8 & 0.56 & 0.95 & 0.74 & -0.14  \\
\hline
\end{tabular}
\end{table}

\subsection{COSA}
\rev{Table~\ref{tab:cosa} presents the performance of COSA for each of the four systems under consideration using the benchmark outlined in section ~\ref{sec:cosa_test_case}. The number of MPI processes used in each of the tests is listed in Table~\ref{tab:cosa_mpi}.} Each benchmark was run three times and the fastest run is presented. For all the benchmarks the variation between the best and worst is less than $2\%$ of the best runtime.

\begin{table}[h]
\caption{\rev{COSA benchmark MPI process counts.}}\label{tab:cosa_mpi}
\begin{tabular}{|c|r|r|r|r|}
\hline
\bf{Nodes} & \makecell{\bf{HPE} \\ \bf{Apollo 70}} & \makecell{\bf{SGI} \\ \bf{ICE XA}} & \bf{Cray XC30} & \bf{Dell EMC} \\
\hline
\hline
4 & 256 & 144 & & \\
8 & 512 & 288 & & 256 \\
16 & 1024 & 576 & & 512 \\
21 & & & 512 &  \\
22 & & & 528 &  \\
32 & 2048 & 1152 & & 1024 \\
40 & & & 960 &\\
\hline
\end{tabular}
\end{table}

\rev{It is important to note that Table~\ref{tab:cosa} outlines the performance as the number of nodes increases. With reference to Table~\ref{tab:cosa_mpi} it can be seen that this equates to different numbers of MPI processes on the different systems. For instance, using 512 MPI processes on the Cray XC30 system gives better performance than on the HPE Apollo 70 system, but at the cost of using significantly more nodes (i.e. 21 nodes vs 8 nodes). If we compare using MPI process counts across the systems it is evident that the performance is generally similar with the SGI ICE XA exhibiting the best performance at lower process counts, although it does not maintain this advantage at higher process counts.}

\rev{However, if we compare on a node to node basis, it is evident that the performance is generally similar between the HPE Apollo 70 and the Dell EMC systems, with the SGI ICE XA and Cray XC30 not providing as quick a time to solution for similar numbers of nodes. Indeed, the Dell EMC and HPE Apollo 70 systems exhibit around $50\%$ increased performance compared to the SGI ICE XA system. The ability to achieve comparable performance using significantly lower numbers of nodes provides potentially lower costs in terms of hardware and energy.}

\begin{table}[h]
\caption{\rev{COSA Correlation values with theoretical peak GFlop/s per node and measured memory bandwidth per node.}}\label{tab:cosa_correl}
\begin{tabular}{|c|r|r|}
\hline
\bf{Nodes} & \bf{Node DP GFlop/s} & \bf{Node memory bandwidth}  \\
\hline
\hline
8 & -0.34 &  -0.86 \\
32 & -0.37 & -0.86  \\
\hline
\end{tabular}
\end{table}

\rev{As with the HPCG benchmarks, we undertook a correlation of COSA performance at 8 and 32 nodes on the Apollo 70, ICE XA, and Dell EMC systems (we did not have results from the Cray XC30 for these node counts) against theoretical floating point performance and measured memory bandwidth. Table~\ref{tab:cosa_correl} shows that COSA has a strong correlation with available memory bandwidth (the negative correlation observed is because we are comparing time to solution numbers to performance metrics, which have an inverse relationship, as time to solution decreases as performance metrics like memory bandwidth increase).}



\begin{table}[t]
\caption{\rev{COSA performance. On the Cray XC30, the test case cannot be run on the same number of nodes as on the other systems, but the results are nevertheless presented as a relative indicator of overall performance. The Dell EMC system does not have enough memory per node to run at fewer than 8 nodes.}}\label{tab:cosa}

\begin{tabular}{|c|r|r|r|r|}
\hline
& \multicolumn{4}{c|}{\makecell{Walltime in seconds \it{(Speedup factor)}}} \\
\hline
\hline
\bf{Nodes} & \makecell{\bf{HPE Apollo } \\ \bf{70 (MPT)}}  & \makecell{\bf{SGI} \\ \bf{ICE XA}} & \bf{Cray XC30} & \bf{Dell EMC} \\
\hline
4 & 2036 \it{( 1.00)}  & 3159 \it{( 1.00)} & &  \\
8 & 1005 \it{( 2.03)} & 1591 \it{( 1.99)} & & 1083 \it{( 1.00)} \\
16 & 484 \it{( 4.21)} & 771 \it{( 4.10)} & & 540 \it{( 2.01)} \\
21 &  &  & 896 \it{( 1.00)} & \\
22 &   &  & 873 \it{( 1.03)} & \\
32 & 239 \it{( 8.52)} & 496 \it{( 6.37)} & & 265 \it{( 4.09)} \\
40 &  &  & 482 \it{( 1.86)} & \\
\hline
\end{tabular}
\end{table}

\subsection{GROMACS}

\rev{Table \ref{tab:gromacs}} shows the performance of GROMACS for each of the systems as a function of number of nodes for the 1400k atom benchmark.


\begin{table*}[t]
\caption{\rev{GROMACS performance results, shown both as the total wallclock time in seconds, together with parallel efficiency, and the application's own performance metric (ns/day), together with the speedup factor for increasing node counts.}}\label{tab:gromacs}
\begin{tabular}{|c|c|c|c|c||c|c|c|c|}
\hline
& \multicolumn{4}{c||}{\makecell{Walltime in seconds \it{(Parallel efficiency \%)}}} & \multicolumn{4}{c|}{\makecell{Performance in ns/day \it{(Speedup factor)}}} \\
\hline
\hline
Nodes & HPE Apollo 70 & Cray XC30 & SGI ICE XA & Dell EMC & HPE Apollo 70 & Cray XC30 & SGI ICE XA & Dell EMC \\
\hline
1 & \makecell[r]{1340.54 \it{(100\%)}} & \makecell[r]{1685.88 \it{(100\%)}} & \makecell[r]{1017.40 \it{(100\%)}} & \makecell[r]{696.54 \it{(100\%)}} & \makecell[r]{1.29 \it{( 1.00)}} & \makecell[r]{1.03 \it{( 1.00)}} & \makecell[r]{1.70 \it{( 1.00)}} & \makecell[r]{2.48 \it{( 1.00)}} \\

2 & \makecell[r]{681.69 \it{( 98\%)}} & \makecell[r]{826.78 \it{(102\%)}} & \makecell[r]{523.95 \it{( 97\%)}} & \makecell[r]{353.66 \it{( 98\%)}} & \makecell[r]{2.54 \it{( 1.97)}} & \makecell[r]{2.09 \it{( 2.04)}} & \makecell[r]{3.30 \it{( 1.94)}} & \makecell[r]{4.89 \it{( 1.97)}} \\

4 & \makecell[r]{345.09 \it{( 97\%)}} & \makecell[r]{452.64 \it{( 93\%)}} & \makecell[r]{275.21 \it{( 92\%)}} & \makecell[r]{183.59 \it{( 95\%)}} & \makecell[r]{5.01 \it{( 3.89)}} & \makecell[r]{3.82 \it{( 3.72)}} & \makecell[r]{6.28 \it{( 3.70)}} & \makecell[r]{9.41 \it{( 3.79)}} \\

8 & \makecell[r]{178.49 \it{( 94\%)}} & \makecell[r]{246.71 \it{( 85\%)}} & \makecell[r]{140.86 \it{( 90\%)}} & \makecell[r]{101.29 \it{( 86\%)}} & \makecell[r]{9.68 \it{( 7.51)}} & \makecell[r]{7.01 \it{( 6.83)}} & \makecell[r]{12.27 \it{( 7.22)}} & \makecell[r]{17.06 \it{( 6.88)}} \\

16 & \makecell[r]{93.50 \it{( 90\%)}} & \makecell[r]{169.33 \it{( 62\%)}} & \makecell[r]{72.55 \it{( 88\%)}} & \makecell[r]{64.76 \it{( 67\%)}} & \makecell[r]{18.48 \it{(14.34)}} & \makecell[r]{10.21 \it{( 9.96)}} & \makecell[r]{23.82 \it{(14.02)}} & \makecell[r]{26.69 \it{(10.76)}} \\

32 & \makecell[r]{62.85 \it{( 67\%)}} & \makecell[r]{96.56 \it{( 55\%)}} & \makecell[r]{69.12 \it{( 46\%)}} & \makecell[r]{35.68 \it{( 61\%)}} & \makecell[r]{27.50 \it{(21.33)}} & \makecell[r]{17.90 \it{(17.46)}} & \makecell[r]{25.00 \it{(14.72)}} & \makecell[r]{48.44 \it{(19.53)}} \\
\hline
\end{tabular}
\end{table*}

\rev{The ordering of the GROMACS performance on the different systems up to 32 compute nodes matches the ordering of compute node performance. It is well known~\cite{GROMACS,ARCHER_bench_perf_report} that GROMACS performance at low process counts for the simulation size is well-correlated with floating point performance so this behaviour is as expected. The majority of MPI time for this GROMACS benchmark is spent in point-to-point communication rather than collective communication. This is underlined by the fact that the scaling behaviour from GROMACS on the Apollo 70 is comparable with that on the other systems, and indeed shows the best scaling performance for the higher node counts. Further investigations is needed to establish what hardware feature(s) are enabling this improved scaling performance, but it could be attributed to the ratio of on-node and off-node MPI communications that the large core counts on the ThunderX2 based systems facilitate.}

\subsection{OpenSBLI}

\rev{Table~\ref{tab:opensbli512}} shows the performance of OpenSBLI for each of the systems as a function of number of nodes for the $512\times512\times512$ benchmark and \rev{Table~\ref{tab:opensbli1024}} shows the performance for the $1024\times1024\times1024$ benchmark.

For the smaller \rev{of the two} benchmark\rev{s}, the Cray XC30 \rev{is the lowest performing in terms of absolute runtime; the SGI ICE XA and HPE Apollo systems are very evenly matched. The Dell system shows the best outright runtime performance. The test case scales well on all systems, achieving 27-28x speedup on 32 nodes (compared to a single node), and the parallel efficiency is at or above 85\%.}

\rev{The larger benchmark requires a minimum of 2 nodes to satisfy its memory requirements (4 nodes on the XC30). It shows very similar performance trends, although the Apollo 70 system now marginally outperforms the SGI ICE XA (on the order of a few \%). As scalability for this test case is even better than for the smaller test case (in fact it is mostly superlinear), the network and communication patterns are evidently not a performance bottleneck here. Taking into account the memory requirements of this test case, as well as the fact that the ThunderX2 CPU does not outperform an Intel Xeon Broadwell CPU on a per-core basis in terms of pure compute performance, the gain for the Apollo70 almost certainly stems from the increased memory bandwidth the Arm-based processor offers.}



\begin{table*}[t]
\caption{\rev{OpenSBLI performance for TGV512ss (100 iterations).}}\label{tab:opensbli512}
\begin{tabular}{|c|c|c|c|c||c|c|c|c|}
\hline
 & \multicolumn{4}{c||}{\makecell{Walltime in seconds \it{(Parallel efficiency \%)}}} & \multicolumn{4}{c|}{\makecell{Performance in iterations/second \it{(Speedup factor)}}} \\
\hline
\hline
Nodes & HPE Apollo 70 & Cray XC30 & SGI ICE XA & Dell EMC & HPE Apollo 70 & Cray XC30 & SGI ICE XA & Dell EMC \\
\hline
1 & 793.65 \it{(100\%)} & 999.91 \it{(100\%)} & 739.89 \it{(100\%)} & 509.92 \it{(100\%)}
& 0.13 \it{( 1.00)} & 0.10 \it{( 1.00)} & 0.14 \it{( 1.00)} & 0.20 \it{( 1.00)} \\

2 & 375.94 \it{(106\%)} & 488.33 \it{(102\%)} & 386.07 \it{( 96\%)} & 257.84 \it{( 99\%)}
& 0.27 \it{( 2.11)} & 0.20 \it{( 2.05)} & 0.26 \it{( 1.68)} & 0.39 \it{( 1.99)} \\

4 & 207.04 \it{( 96\%)} & 288.80 \it{( 87\%)} & 218.82 \it{( 85\%)} & 131.90 \it{( 97\%)}
& 0.48 \it{( 3.83)} & 0.35 \it{( 3.46)} & 0.46 \it{( 3.29)} & 0.76 \it{( 3.88)} \\

8 & 105.60 \it{( 94\%)} & 142.35 \it{( 88\%)} & 101.65 \it{( 91\%)} & 68.05 \it{( 94\%)}
& 0.95 \it{( 7.51)} & 0.70 \it{( 7.02)} & 0.98 \it{( 7.00)} &1.47 \it{( 7.50)} \\

16 & 53.05 \it{( 94\%)} & 65.85 \it{( 95\%)} & 51.93 \it{( 89\%)} & 34.26 \it{( 93\%)}
& 1.89 \it{(14.96)} & 1.52 \it{(15.18)} & 1.93 \it{(13.79)} & 2.92 \it{(14.89)} \\

32 & 28.59 \it{( 87\%)} & 35.09 \it{( 89\%)} & 26.75 \it{( 86\%)} & 18.64 \it{( 85\%)}
& 3.50 \it{(27.76)} & 2.85 \it{(28.50)} & 3.74 \it{(26.71)} & 5.36 \it{(27.33)} \\
\hline
\end{tabular}
\end{table*}

\begin{table*}[t]
\caption{\rev{OpenSBLI performance for TGV1024ss (10 iterations). The memory requirements of this test case are such that a minimum of 2 nodes are needed for all systems (apart from the Cray XC30, which needs 4 nodes). The efficiency and speedup factors are provided relative to the lowest possible node count for each system.}}\label{tab:opensbli1024}
\begin{tabular}{|c|r|r|r|r||r|r|r|r|}
\hline
 & \multicolumn{4}{|c||}{\makecell{Walltime in seconds \it{(Parallel efficiency \%)}}} & \multicolumn{4}{c|}{\makecell{Performance in iterations/second \it{(Speedup factor)}}} \\
\hline
\hline
Nodes & HPE Apollo 70 & Cray XC30 & SGI ICE XA & Dell EMC & HPE Apollo 70 & Cray XC30 & SGI ICE XA & Dell EMC \\
\hline
2 & 312.50 \it{(100\%)} &  & 309.33 \it{(100\%)} & 213.47 \it{(100\%)}
& 0.03 \it{( 1.00)} &  & 0.03 \it{( 1.00)} & 0.05 \it{( 1.00)} \\

4 & 158.73 \it{( 98\%)} & 301.07 \it{(100\%)} & 174.44 \it{( 89\%)} & 105.91 \it{(101\%)}
& 0.06 \it{( 1.97)} & 0.03 \it{( 1.00)} & 0.06 \it{( 2.00)} & 0.09 \it{( 2.02)} \\

8 & 80.00 \it{( 98\%)} & 104.95 \it{(143\%)} & 75.88 \it{(102\%)} &  53.14 \it{(100\%)}
& 0.12 \it{( 3.91)} & 0.09 \it{( 2.87)} & 0.13 \it{( 4.33)} & 0.19 \it{( 4.00)} \\

16 & 30.02 \it{(130\%)} &  49.10 \it{(153\%)} &  39.05 \it{( 99\%)} & 25.05 \it{(107\%)}
& 0.33 \it{(10.41)} & 0.20 \it{( 6.13)} & 0.26 \it{( 8.67)} & 0.40 \it{( 8.52)} \\

32 & 20.80 \it{( 94\%)} & 28.27 \it{(133\%)} & 22.19 \it{( 87\%)} & 13.14 \it{(102\%)}
& 0.48 \it{(15.02)} & 0.35 \it{(10.65)} & 0.45 \it{(15.00)} & 0.76 \it{(16.25)} \\
\hline
\end{tabular}
\end{table*}

\subsection{Nektar++}


Table~\ref{tab:nektar} shows the performance of Nektar++ running on both the SGI ICE XA and HPE Apollo 70 systems, represented as runtime to solution, and showing \rev{both the parallel efficiency and }the speedup relative to one node. Comparing the performance on a per-node basis, the Apollo 70 system displays better performance than the SGI ICE XA at low node counts, however the gap closes at 8 nodes \rev{and the Arm-based system falls behind. One potential explanation is that the test-case does not provide sufficient work which limits the performance with higher core-count on the ARM based system. Unfortunately, a larger test-case was not available at the time of writing. It is also evident that the ARM compiler is providing performance benefits at two to eight nodes when compared to the GNU compiler. This, coupled with the super-linear speed up in those regions, suggests that the ARM compiler is better utilising the cache compared to the GNU compiler.}

\begin{table}[t]
\caption{\rev{Nektar++ performance.}}\label{tab:nektar}
\begin{tabular}{|c|c|c|c|}
\hline
 & \multicolumn{3}{c|}{\makecell[c]{Walltime in seconds  \\ \it{(Parallel Efficiency \& Speedup factor)}}} \\
\hline
\hline
\makecell[c]{Nodes\\ } & \makecell[c]{HPE Apollo 70\\(GNU+OpenMPI)} & \makecell[c]{HPE Apollo 70\\(ARM+OpenMPI)} & \makecell[c]{SGI ICE XA\\ } \\
\hline
1 & \makecell[c]{1751 \\ \it{(100\% \&     1.00)}}  & \makecell[c]{1800 \\ \it{(100\% \&  1.00)}} & \makecell[c]{2621 \\ \it{(100\% \&   1.00)}}\\

2 & \makecell[c]{1362 \\ \it{(  64\% \&     1.29)}} & \makecell[c]{842 \\ \it{(107\% \&  2.14)}} & \makecell[c]{1227 \\ \it{(107\% \&   2.13)}}\\

4 & \makecell[c]{466 \\ \it{(  94\% \&     3.76)}} & \makecell[c]{364 \\ \it{(124\%\&  4.94)}} & \makecell[c]{502 \\ \it{(131\%  \& 5.22)}}\\

8 & \makecell[c]{185 \\ \it{(118\% \& 9.46)}} & \makecell[c]{172 \\ \it{(122\% \& 10.46)}} & \makecell[c]{159 \\ \it{(206\% \&  16.48)}}\\

16 & \makecell[c]{109 \\ \it{(100\% \& 16.06)}} & \makecell[c]{109 \\ \it{(103\% \& 16.51)}} & \makecell[c]{88 \\ \it{(186\% \&  20.45)} }\\
\hline
\end{tabular}
\end{table}

\subsection{Application Portability and Maturity of the Software Stack}
The HPE Apollo 70 system offers a range of compilers, including the GNU and Arm compiler suites, supporting C, C++, and Fortran applications. These are complemented by the Arm performance libraries~\cite{Arm_Perf_Libs}, which provide optimised implementations of BLAS, LAPACK, FFT and standard maths routines. There are also a number of different MPI implementations available for use on the system, including OpenMPI and MPICH, with MVAPICH2 shortly to be available. The availability of this combination of libraries and compilers has made porting the applications we have benchmarked for this paper straightforward, requiring no code modification, and only simple adaptations of the build processes for the applications.

The software stack that is available on the Apollo 70 is mature and no major problems were encountered when building the applications or evaluating the benchmarks. The only performance issue that we discovered is the disappointing performance of the MPI collective operations, and this is under active investigation. All in all, we found the software stack to be complete and sufficiently mature to be able to build and run complex real-life applications without difficulties. \rev{It is worth noting however that the Apollo 70 system used for the experiments in this paper does not have access to a high performance parallel file system yet (this part of the system is still under development), and the benchmarks therefore do not include any IO; as such we cannot assess that aspect of the system. 

Finally, although we did not encounter any significant issues when porting applications to the Arm-based system, the same is not true for performance profiling, in particular with regards to memory access and usage. The tools that are available on the Apollo 70 focus largely on CPU and network performance. However, as discussed above, the additional memory channels available on the Apollo 70 are key to delivering improved performance for applications that are limited by memory bandwidth, but there are no profiling tools available on the system that would allow us to quantify the effect directly. This is a limitation that should be addressed in order to allow application developers to fully understand the performance of their codes in the Arm ecosystem.}

\section{Conclusions}
\begin{figure}[h]
\centering
\includegraphics[width=\linewidth]{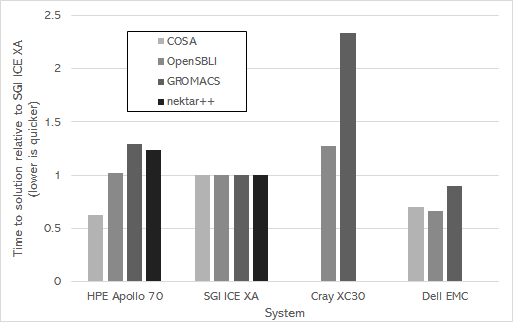}
\caption{Comparison of application time to solution using 16 nodes normalised to the SGI ICE XA system. Values lower than one are faster than the SGI ICE XA system.}
\label{fig:totalgraph}
\end{figure}

This paper gives a first overview of the performance of scientific applications running at scale on a production HPC system that is based on Arm processors. Our experience gained from porting a wide range of applications to the ThunderX2 processors and using the Arm supported software ecosystem demonstrates that this is possible with minimal effort. We have also demonstrated that applications can achieve similar, or better, performance on such a system when compared with a range of existing HPC system architectures, \rev{as highlighted by Figure~\ref{fig:totalgraph}}. Due to the additional memory channels available to them, the ThunderX2 processors provide clear benefits for applications that are memory bandwidth bound. However, when comparing on a per-node basis, the Apollo 70 also exhibits good performance on applications that are compute rather than memory intensive. Our conclusion is that the Arm ecosystem, both the hardware and the softwar stack, is already proving itself to be a viable alternative to the status quo.

\section{Acknowledgements}
The Fulhame HPE Apollo 70 system is supplied to EPCC as part of the Catalyst UK programme, a collaboration with Hewlett Packard Enterprise, Arm and SUSE to accelerate the adoption of Arm based supercomputer applications in the UK.

This work used the Cirrus UK National Tier-2 HPC Service at EPCC (http://www.cirrus.ac.uk) funded by the University of Edinburgh and EPSRC (EP/P020267/1).

This work used the ARCHER UK National Supercomputing Service (http://www.archer.ac.uk).

We thank the University of Cambridge for access to resources provided by the Cambridge Service for Data Driven Discovery (CSD3). CSD3 is operated by the University of Cambridge Research Computing Service (http://www.csd3.cam.ac.uk/), provided by Dell EMC and Intel using Tier-2 funding from the Engineering and Physical Sciences Research Council (capital grant EP/P020259/1), and DiRAC funding from the Science and Technology Facilities Council (www.dirac.ac.uk).

\bibliographystyle{ACM-Reference-Format}
\bibliography{arm-perf-pasc19}


\begin{thebibliography}{18}


\ifx \showCODEN    \undefined \def \showCODEN     #1{\unskip}     \fi
\ifx \showDOI      \undefined \def \showDOI       #1{#1}\fi
\ifx \showISBNx    \undefined \def \showISBNx     #1{\unskip}     \fi
\ifx \showISBNxiii \undefined \def \showISBNxiii  #1{\unskip}     \fi
\ifx \showISSN     \undefined \def \showISSN      #1{\unskip}     \fi
\ifx \showLCCN     \undefined \def \showLCCN      #1{\unskip}     \fi
\ifx \shownote     \undefined \def \shownote      #1{#1}          \fi
\ifx \showarticletitle \undefined \def \showarticletitle #1{#1}   \fi
\ifx \showURL      \undefined \def \showURL       {\relax}        \fi
\providecommand\bibfield[2]{#2}
\providecommand\bibinfo[2]{#2}
\providecommand\natexlab[1]{#1}
\providecommand\showeprint[2][]{arXiv:#2}

\bibitem[\protect\citeauthoryear{??}{ARC}{[n. d.]a}]%
        {ARCHER_Benchmarks}
 \bibinfo{year}{[n. d.]}\natexlab{a}.
\newblock \bibinfo{title}{ARCHER Benchmarks}.
\newblock
\newblock
\urldef\tempurl%
\url{http://www.github.com/hpc-uk/archer-benchmarks}
\showURL{%
Retrieved 10 January 2019 from \tempurl}


\bibitem[\protect\citeauthoryear{??}{ARC}{[n. d.]b}]%
        {ARCHER_Usage}
 \bibinfo{year}{[n. d.]}\natexlab{b}.
\newblock \bibinfo{title}{ARCHER usage data}.
\newblock
\newblock
\urldef\tempurl%
\url{https://www.archer.ac.uk/status/codes}
\showURL{%
Retrieved 10 January 2019 from \tempurl}


\bibitem[\protect\citeauthoryear{??}{Arm}{[n. d.]}]%
        {Arm_Perf_Libs}
 \bibinfo{year}{[n. d.]}\natexlab{}.
\newblock \bibinfo{title}{Arm Performance Libraries}.
\newblock
\newblock
\urldef\tempurl%
\url{https://developer.arm.com/products/software-development-tools/hpc/arm-performance-librariess}
\showURL{%
Retrieved 15 January 2019 from \tempurl}


\bibitem[\protect\citeauthoryear{??}{HEC}{[n. d.]}]%
        {HECBioSim_benchmarks}
 \bibinfo{year}{[n. d.]}\natexlab{}.
\newblock \bibinfo{title}{HEC BioSim Benchmarks}.
\newblock
\newblock
\urldef\tempurl%
\url{http://www.hecbiosim.ac.uk/benchmarks}
\showURL{%
Retrieved 11 January 2019 from \tempurl}


\bibitem[\protect\citeauthoryear{Abraham, Murtola, Schulz, P\'{a}ll, Smith,
  Hess, and Lindahl}{Abraham et~al\mbox{.}}{2015}]%
        {GROMACS}
\bibfield{author}{\bibinfo{person}{Mark~James Abraham}, \bibinfo{person}{Teemu
  Murtola}, \bibinfo{person}{Roland Schulz}, \bibinfo{person}{Szil\'{a}rd
  P\'{a}ll}, \bibinfo{person}{Jeremy~C. Smith}, \bibinfo{person}{Berk Hess},
  {and} \bibinfo{person}{Erik Lindahl}.} \bibinfo{year}{2015}\natexlab{}.
\newblock \showarticletitle{GROMACS: High performance molecular simulations
  through multi-level parallelism from laptops to supercomputers}.
\newblock \bibinfo{journal}{\emph{SoftwareX}}  \bibinfo{volume}{1-2}
  (\bibinfo{year}{2015}), \bibinfo{pages}{19 -- 25}.
\newblock
\showISSN{2352-7110}
\urldef\tempurl%
\url{https://doi.org/10.1016/j.softx.2015.06.001}
\showDOI{\tempurl}


\bibitem[\protect\citeauthoryear{Bailey, Barton, Lasinski, and Simon}{Bailey
  et~al\mbox{.}}{1993}]%
        {NASBenchmark}
\bibfield{author}{\bibinfo{person}{David Bailey}, \bibinfo{person}{John
  Barton}, \bibinfo{person}{Thomas Lasinski}, {and} \bibinfo{person}{Horst
  Simon}.} \bibinfo{year}{1993}\natexlab{}.
\newblock \showarticletitle{The NAS parallel benchmarks}.
\newblock  (\bibinfo{date}{08} \bibinfo{year}{1993}).
\newblock
\urldef\tempurl%
\url{https://doi.org/10.2172/983318}
\showDOI{\tempurl}


\bibitem[\protect\citeauthoryear{Blanc}{Blanc}{[n. d.]}]%
        {Arm_energy}
\bibfield{author}{\bibinfo{person}{Mont Blanc}.} \bibinfo{year}{[n.
  d.]}\natexlab{}.
\newblock \bibinfo{title}{D9.6 Performance analysis of applications and
  mini-applications and benchmarking on the project test platforms}.
\newblock
\newblock
\urldef\tempurl%
\url{https://www.montblanc-project.eu/project/deliverables}
\showURL{%
Retrieved March 2019 from \tempurl}


\bibitem[\protect\citeauthoryear{Cantwell, Moxey, Comerford, Bolis, Rocco,
  Mengaldo, Grazia, Yakovlev, Lombard, and et~al.}{Cantwell
  et~al\mbox{.}}{2015}]%
        {NektarPaper}
\bibfield{author}{\bibinfo{person}{CD Cantwell}, \bibinfo{person}{D Moxey},
  \bibinfo{person}{A Comerford}, \bibinfo{person}{A Bolis}, \bibinfo{person}{G
  Rocco}, \bibinfo{person}{G Mengaldo}, \bibinfo{person}{D~De Grazia},
  \bibinfo{person}{S Yakovlev}, \bibinfo{person}{J-E Lombard}, {and}
  \bibinfo{person}{D~Ekelschot et al.}} \bibinfo{year}{2015}\natexlab{}.
\newblock \showarticletitle{Nektar++: An open-source spectral/hp element
  framework}.
\newblock \bibinfo{journal}{\emph{Computer Physics Communications}}
  \bibinfo{volume}{192} (\bibinfo{year}{2015}), \bibinfo{pages}{205--219}.
\newblock


\bibitem[\protect\citeauthoryear{Dongarra and Luszczek}{Dongarra and
  Luszczek}{2004}]%
        {HPCChallenge}
\bibfield{author}{\bibinfo{person}{Jack Dongarra} {and} \bibinfo{person}{Piotr
  Luszczek}.} \bibinfo{year}{2004}\natexlab{}.
\newblock \showarticletitle{Introduction to the HPCChallenge Benchmark Suite}.
\newblock  (\bibinfo{date}{12} \bibinfo{year}{2004}).
\newblock


\bibitem[\protect\citeauthoryear{Dongarra, Heroux, and Luszczek}{Dongarra
  et~al\mbox{.}}{2015}]%
        {HPCG}
\bibfield{author}{\bibinfo{person}{Jack~J. Dongarra},
  \bibinfo{person}{Michael~A. Heroux}, {and} \bibinfo{person}{Piotr Luszczek}.}
  \bibinfo{year}{2015}\natexlab{}.
\newblock \showarticletitle{HPCG Benchmark : a New Metric for Ranking High
  Performance Computing Systems ∗}.
\newblock


\bibitem[\protect\citeauthoryear{Dongarra, Luszczek, and Petitet}{Dongarra
  et~al\mbox{.}}{2003}]%
        {HPL}
\bibfield{author}{\bibinfo{person}{Jack~J. Dongarra}, \bibinfo{person}{Piotr
  Luszczek}, {and} \bibinfo{person}{Antoine Petitet}.}
  \bibinfo{year}{2003}\natexlab{}.
\newblock \showarticletitle{The LINPACK Benchmark: past, present and future}.
\newblock \bibinfo{journal}{\emph{Concurrency and Computation: Practice and
  Experience}}  \bibinfo{volume}{15} (\bibinfo{year}{2003}),
  \bibinfo{pages}{803--820}.
\newblock


\bibitem[\protect\citeauthoryear{Forum}{Forum}{[n. d.]}]%
        {MPI}
\bibfield{author}{\bibinfo{person}{MPI Forum}.} \bibinfo{year}{[n.
  d.]}\natexlab{}.
\newblock \bibinfo{title}{MPI: A Message-Passing Interface Standard, Version
  2.2}.
\newblock
\newblock
\urldef\tempurl%
\url{http://www.mpi-forum.org}
\showURL{%
Retrieved September 2009 from \tempurl}


\bibitem[\protect\citeauthoryear{Jackson and Campobasso}{Jackson and
  Campobasso}{2011}]%
        {COSA_CSRD}
\bibfield{author}{\bibinfo{person}{Adrian Jackson} {and} \bibinfo{person}{{M.
  Sergio} Campobasso}.} \bibinfo{year}{2011}\natexlab{}.
\newblock \showarticletitle{Shared-memory, distributed-memory, and mixed-mode
  parallelisation of a CFD simulation code}.
\newblock \bibinfo{journal}{\emph{Computer Science - Research and Development}}
  \bibinfo{volume}{26}, \bibinfo{number}{3-4} (\bibinfo{date}{6}
  \bibinfo{year}{2011}), \bibinfo{pages}{187--195}.
\newblock
\showISSN{1865-2034}
\urldef\tempurl%
\url{https://doi.org/10.1007/s00450-011-0162-4}
\showDOI{\tempurl}


\bibitem[\protect\citeauthoryear{Jackson, Campobasso, and Drofelnik}{Jackson
  et~al\mbox{.}}{2018}]%
        {COSA_CF}
\bibfield{author}{\bibinfo{person}{William Jackson}, \bibinfo{person}{{M.
  Sergio} Campobasso}, {and} \bibinfo{person}{Jernej Drofelnik}.}
  \bibinfo{year}{2018}\natexlab{}.
\newblock \showarticletitle{Load balance and Parallel I/O: Optimising COSA for
  large simulations}.
\newblock \bibinfo{journal}{\emph{Computers and Fluids}} (\bibinfo{date}{9 3}
  \bibinfo{year}{2018}).
\newblock
\showISSN{0045-7930}
\urldef\tempurl%
\url{https://doi.org/10.1016/j.compfluid.2018.03.007}
\showDOI{\tempurl}


\bibitem[\protect\citeauthoryear{Jacobs, Jammy, and Sandham}{Jacobs
  et~al\mbox{.}}{2017}]%
        {OSBLI_2017}
\bibfield{author}{\bibinfo{person}{C.~T. Jacobs}, \bibinfo{person}{S.~P.
  Jammy}, {and} \bibinfo{person}{N.~D. Sandham}.}
  \bibinfo{year}{2017}\natexlab{}.
\newblock \showarticletitle{{OpenSBLI: A framework for the automated derivation
  and parallel execution of finite difference solvers on a range of computer
  architectures}}.
\newblock \bibinfo{journal}{\emph{{Journal of Computational Science}}}
  \bibinfo{volume}{18} (\bibinfo{year}{2017}), \bibinfo{pages}{12--23}.
\newblock
\urldef\tempurl%
\url{https://doi.org/10.1016/j.jocs.2016.11.001}
\showDOI{\tempurl}


\bibitem[\protect\citeauthoryear{McIntosh-Smith, Price, Deakin, and
  Poenaru}{McIntosh-Smith et~al\mbox{.}}{2018}]%
        {Bristol_CUG18}
\bibfield{author}{\bibinfo{person}{Simon McIntosh-Smith},
  \bibinfo{person}{James Price}, \bibinfo{person}{Tom Deakin}, {and}
  \bibinfo{person}{Andrei Poenaru}.} \bibinfo{year}{2018}\natexlab{}.
\newblock \showarticletitle{Comparative Benchmarking of the First Generation of
  HPC-Optimised Arm Processors on Isambard}. In \bibinfo{booktitle}{\emph{Cray
  User Group}}.
\newblock


\bibitem[\protect\citeauthoryear{Müller, van Waveren, Lieberman, Whitney,
  Saito, Kumaran, Baron, Brantley, Parrott, Elken, Feng, and Ponder}{Müller
  et~al\mbox{.}}{2010}]%
        {SPECMPI}
\bibfield{author}{\bibinfo{person}{Matthias Müller}, \bibinfo{person}{Matthijs
  van Waveren}, \bibinfo{person}{Ron Lieberman}, \bibinfo{person}{Brian
  Whitney}, \bibinfo{person}{Hideki Saito}, \bibinfo{person}{Kalyan Kumaran},
  \bibinfo{person}{John Baron}, \bibinfo{person}{William Brantley},
  \bibinfo{person}{Chris Parrott}, \bibinfo{person}{Tom Elken},
  \bibinfo{person}{Huiyu Feng}, {and} \bibinfo{person}{Carl Ponder}.}
  \bibinfo{year}{2010}\natexlab{}.
\newblock \showarticletitle{SPEC MPI2007-an application benchmark suite for
  parallel systems using MPI}.
\newblock \bibinfo{journal}{\emph{Concurrency and Computation: Practice and
  Experience}}  \bibinfo{volume}{22} (\bibinfo{date}{02} \bibinfo{year}{2010}),
  \bibinfo{pages}{191--205}.
\newblock
\urldef\tempurl%
\url{https://doi.org/10.1002/cpe.1535}
\showDOI{\tempurl}


\bibitem[\protect\citeauthoryear{Turner and Salmond}{Turner and
  Salmond}{2018}]%
        {ARCHER_bench_perf_report}
\bibfield{author}{\bibinfo{person}{Andrew Turner} {and}
  \bibinfo{person}{Jeffrey Salmond}.} \bibinfo{year}{2018}\natexlab{}.
\newblock \bibinfo{title}{{hpc-uk/archer-benchmarks: Initial performance
  comparison report}}.
\newblock
\newblock
\urldef\tempurl%
\url{https://doi.org/10.5281/zenodo.1288378}
\showDOI{\tempurl}


\end{thebibliography}

\end{document}